\documentclass[prl,aps,twocolumn,groupedaddress,floats,showpacs,final]{revtex4-1}
\usepackage{graphicx}
\usepackage{dcolumn}
\usepackage{bm}
\usepackage{color}
\usepackage[caption=false]{subfig}
\definecolor{DarkGreen}{rgb}{0,0.7,0.08} 
\definecolor{Grey}{rgb}{0.5,0.5,0.5}
\definecolor{Red}{rgb}{0.8,0.3,0.3}
\definecolor{Green}{rgb}{0.1,0.8,0.1}
\definecolor{Blue}{rgb}{0.1,0.1,0.8}
\newcommand{\oldtext}[1]{}

\begin{document}

\title{Demixing in symmetric supersolid mixtures}
\author{Piyush Jain$^{1,3}$,  Saverio Moroni$^2$,  Massimo Boninsegni$^{1,2}$ and Lode Pollet$^4$}
\affiliation{$^1$ Department of Physics, University of Alberta, Edmonton, Alberta, Canada}
\affiliation{$^2$ {SISSA Scuola Internazionale Superiore di Studi Avanzati and DEMOCRITOS National 
Simulation Center,
Istituto Officina dei Materiali del CNR Via Bonomea 265, I-34136, Trieste, Italy}}
\affiliation{$^3$ Department of Physics and Astronomy, University of British Columbia, Vancouver, British Columbia, Canada}
\affiliation{$^4$ Department of Physics, Arnold Sommerfeld Center for Theoretical Physics and Center for NanoScience, University of Munich, Theresienstrasse 37, 80333 Munich, Germany}
\date{\today}                                           % Activate to display a given date or no date

%-------------------------------------------------------------------------------------------------------------
\begin{abstract}

The droplet crystal phase of a symmetric binary mixture of Rydberg-blockaded dipolar Bose gases is studied by computer simulation.  At high temperature each droplet comprises on average equal numbers of particles of either component, but the two components demix below the supersolid transition temperature, {\it i.e.}, droplets mostly consist of particles of one component. Droplets consisting of the same component will also favor clustering. Demixing is driven by quantum  tunnelling of particles across droplets over the system, and does not take place in a non-superfluid crystal.  This effect should be easily detectable in a cold gas experiment.

\end{abstract}

\maketitle

%-------------------------------------------------------------------------------------------------------------
The separation at low temperature of the individual components of a mixture (demixing) is a topic of long standing interest in physics and chemistry. It  typically occurs due to differences in  interactions,  but quantum-mechanical effects, either related to mass differences  (zero-point motion) or to quantum statistics, can play an important role.  One of the most interesting examples of binary mixture whose phase diagram is significantly affected by quantum statistics, is that of a mixture of the two isotopes of helium, which historically had a profound influence on refrigerating \cite{bbp}. 
\\ \indent
In recent years, renewed interest in quantal binary mixtures has been motivated by experimental advances in cold atom physics, allowing for the controlled   study of the phase diagram of multicomponent systems, notably mixtures of Bose-Einstein condensates (BECs). The potential advantage that ultracold gases offer, compared to ordinary condensed matter systems, is that demixing can be  observed  on considerably shorter time scales and directly in real space with individual particle resolution \cite{Schauss2012}.  
The first experimental realization of a two-component Bose mixture, consisting of two hyperfine states of Rb,  was reported in 1997 \cite{myatt1997}; successively, a mixture of  different atomic species (K and Rb) was also stabilized \cite{Modugno2001}. 
\\ \indent
Early work on bosonic mixtures predicted phase separation at exactly zero temperature for isotopes of different masses or concentrations  \cite{Chester1955, Miller1978}. The focus of more recent theoretical work has been on identifying the conditions under which demixing occurs in binary BEC mixtures with repulsive interactions \cite{Ho1996, Ao1998, Timmermans1998, Pu1998, Esry1999, Trippenbach2000,Shi2000,Ma2006, Alon2006, Sakhel2008}. The systems we are looking at in this work have a SU(2)-symmetric Hamiltonian
\begin{eqnarray}
H &= & \int d \bm{r} \psi^{\dagger}_{\sigma}(\bm{r}) H_0 \psi_{\sigma}(\bm{r})  \label{eq:ham}\\
{} & + & \frac{1}{2} \int d \bm{r} \int d\bm{r'} 
V(\bm{r} - \bm{r'})\psi^{\dagger}_{\sigma}(\bm{r}) \psi^{\dagger}_{\sigma'}(\bm{r'})\psi_{\sigma'}(\bm{r'})\psi_{\sigma}(\bm{r}), \nonumber
\end{eqnarray}
where $H_0$ is the kinetic energy (it may contain also an external one-body potential), $V(\bm{r} - \bm{r'})$ the two-body interaction term, $\psi_{\sigma}$ the Bose field operators {for components $\sigma = 1,2$} and the summation is performed over repeated Greek indices. The dispersion for both species is identical.
Separation of species 1 and 2  is usually characterized in terms of a parameter $\Delta = U_{11} U_{22} - U_{12}^2$, defined in terms of the relative intraspecies ($U_{11}, U_{22}$) and interspecies  ($U_{12}$) characteristic interaction strengths. When $\Delta < 0$,  {i.e.}, unlike particles repel more strongly, separation of the two components occurs, whereas for $\Delta \geq 0$ they  remain mixed \cite{Ho1996, Ao1998, Timmermans1998,Pu1998, Esry1999, Trippenbach2000,Shi2000, Ma2006, Sakhel2008,Papp2008}. {However}, in trapped bosonic mixtures, it was recently shown \cite{jain2010} that the effective attraction between identical Bose particles leads to demixing at low finite temperatures, even in purely repulsive (dipolar) Bose mixtures that are {SU(2) {\it symmetric}  ({i.e.}, a stricter condition than $\Delta=0$). 
\\ \indent
Recent studies of demixing arising from quantum-mechanical effects have focused primarily on the gas or liquid phases, although a solid $^3$He-$^4$He mixture, extensively studied experimentally and theoretically in the past \cite{mullin}, remains arguably the most important example of low temperature isotopic separation. In that case,  all interactions are very nearly identical, and effects of quantum statistics are negligible at the temperature at which onset of phase separation is observed. Separation is driven by single-particle tunnelling,  and can be ascribed to the mass difference between the two isotopes, resulting in different equilibrium densities. 
Cold gases seem the ideal playground in which fundamental questions about demixing in the solid phase can be addressed experimentally, with a high degree of control. The observable selection of ground states has been suggested for isotopic SU(2) symmetric BECs~\cite{Kuklov2002}, when lowering the temperature below the superfluid transition. Potentially interesting is the study of demixing  in solid phases
not occurring in ordinary condensed matter \cite{emphasis}. 
\\ \indent
In this Letter, we predict low-temperature demixing in  a  SU(2) symmetric two-component  {\it supersolid} Bose mixture.  The crystalline phase considered here features a multiply-occupied unit cell (essentially a ``droplet" or cluster of particles), and turns supersolid at sufficiently low temperature through the tunnelling of particles between adjacent (locally superfluid) droplets. It arises at low temperature in the presence of a specific  pair-wise interaction, featuring a soft repulsive core at short distances \cite{notegross,Cinti2010,Saccani1,Boninsegnib}. This type of interaction can be artifically fashioned by means of a mechanism known as the Rydberg Blockade \cite{Lukin2001}. The excitation spectrum of the one-component system was recently shown to display two distinct modes in the supersolid phase \cite{spectrum}.
\\ \indent
{Demixing} in the two-component system is a consequence of the same physical mechanism that underlies the supersolid transition 
in the single-component system. In the normal solid phase, at temperatures where quantum statistical effects are negligible, unit cells feature on average the same numbers of particles of either species. At lower temperatures, but where the system is still insulating, unit cells comprise prevalently particles of one of the two species due to exchanges within each droplet. Further, below the supersolid transition, macroscopic separation (demixing) of the two components is also predicted, {\it i.e.}, clusters containing particles of a given type are surrounded by clusters of the same type of particles. 
This effect should be observable experimentally, for example using a bosonic system where 
the bosons have an internal degree of freedom. 
%system of cold atoms of spin $S=1$.
\\ \indent
Our model consists of a two-component  mixture,  with particles confined to moving in two dimensions and with only pair-wise interactions. Let 1 and 2 be the two components, both obeying Bose statistics, and $N_1$ and $N_2$ the corresponding numbers of particles. 
%We assume that the mixture is symmetric, i.e., the two components have the  same mass ($m$) and that all interactions are identical. 
We are interested here in the type of soft-core repulsive potentials which are known to underlie the supersolid phase described above \cite{Boninsegnib}, 
and for definiteness we take  the same expression considered in Ref. \cite{Cinti2010}, namely
\begin{eqnarray}
\label{pot1}
V(r) = \left\{
\begin{array}{rl} C/a^3 & \text{if } r \le a\, \\
\\
C/r^3 & \text{if } r> a\,
\end{array}
\right., %\; \; \; v_b(r) = \frac{D}{r^3 + R_c^3} \; .
\end{eqnarray}
with $a$ being the interaction cutoff. 
Henceforth, all lengths are expressed in terms of the characteristic unit $r_\circ\equiv mC/\hbar^2$, and we also introduce the energy scale $\epsilon_\circ=C/r_\circ^3=\hbar^2/mr_\circ^2$.  It is important to state upfront that the main physical results discussed here do not depend on the detailed form of the potential for $r>a$, but only on the existence of a soft repulsive core.
%MB the results of that other study also don't depend on the specific form of potential --
%MB it is a different system altogether, no need to make this comment I don't think
%unlike the system studied in Ref. \cite{jain2010}.
%MB
The Hamiltonian is symmetric with respect to an interchange of the component labels 1 and 2. Thus, any phase separation must arise exclusively as a result of Bose statistics. 
%MB This statement reads out of context here, if we want to keep it we need to move it somewhere else
%Recently, spatially ordered structures in a two-dimensional Rydberg gas have been observed %experimentally~\cite{Schauss2012}.  
%MB
\\ \indent
With the two-body interaction (\ref{pot1}) we have investigated the finite temperature equilibrium properties of the system using Quantum Monte Carlo simulations based on the continuous-space Worm algorithm \cite{Boninsegni2006,worm2} in the grand canonical ensemble. This methodology yields results that are numerically exact to within a statistical error, which may be made arbitrarily small by sampling from a sufficient number of configurations of the system. We consider a system with a Rydberg blockade interaction cutoff $a/r_{\circ} = 0.27$, contained in a square simulation cell with periodic boundary conditions in both directions; we adjust the chemical potential so that the mean interparticle separation is $r_s = 0.15$ (equivalent to an average density $\rho r_{\circ}^2 = 45$). For this choice of parameters, the one-component system displays a low temperature supersolid phase \cite{Cinti2010}.

\begin{figure}[!t]
\begin{center}
\begin{minipage}{\columnwidth}
  \centering
    %\captionof*{figure}{(a)}\vspace{-2mm}
    \hspace{1cm}(a)\\
  \includegraphics[scale=1]{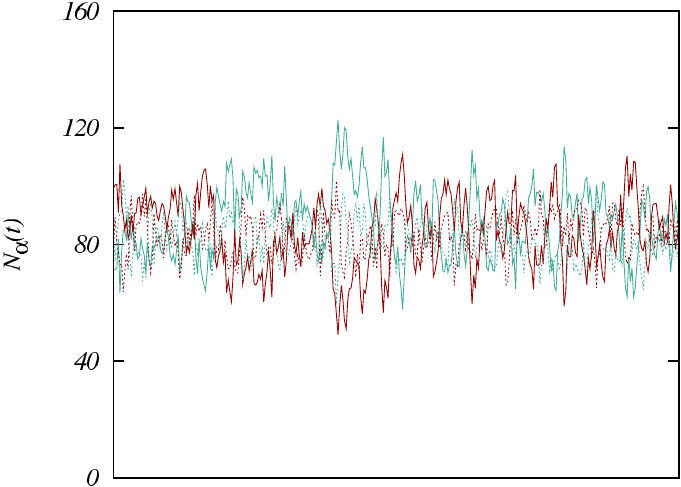}\\ \vspace{2mm}
  \label{fig:test1}
    %\captionof*{figure}{(b)}\vspace{-2mm}
    \hspace{1cm}(b)\\ 
 \includegraphics[scale=1]{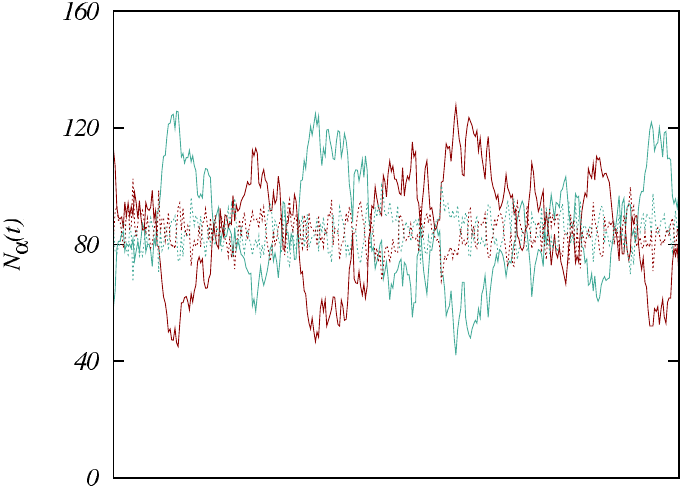}\\
 \vspace{2mm}
  \label{fig:test2}
    %\captionof*{figure}{(c)}\vspace{-2mm}
    \hspace{1cm}(c)\\
  \includegraphics[scale=1]{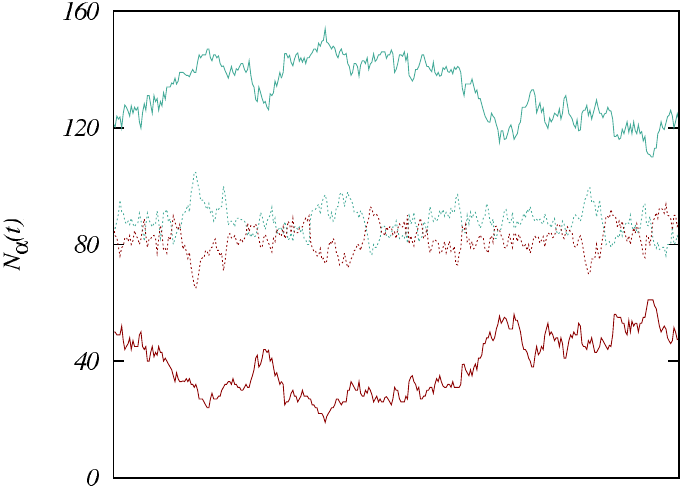}\\
  \vspace{2mm}
  \label{fig:test3}
  %\caption*{\figurename s:~The general caption}
  \caption[justification=justified]{(Color online) Number of particles per species $N_{\alpha}$ ($\alpha = 1,2$) as a function of simulation time at temperatures (a) $T=20 \epsilon_\circ$, (b) $T=10 \epsilon_\circ$ and (c) $T=5 \epsilon_\circ$ (all temperatures are in units of $\epsilon_\circ$). Phase equilibrium requires that the chemical potentials for both species be equal, with $\mu_1 = \mu_2$, chosen so the total number of particles is $N \approx 170$. Note the blue curve corresponds to species $\alpha = 1$, the red to $\alpha = 2$ whereas the dotted curves represent the equivalent simulation but in the case of Boltzmann statistics.  }\vspace{-5mm}
\label{fig:gcresults}
\end{minipage}%
\end{center}
\end{figure}

In our grand canonical simulations, we can directly observe the phenomenon of phase separation, which manifests itself as spontaneous symmetry breaking between the populations of species 1 and 2. That is, at equilibrium either species will have a greater population, although the total population of both species is fixed by the chemical potential. This effect is shown clearly in Fig.~\ref{fig:gcresults} which plots the number of particles of each species as a function of simulation time at three temperatures. Obviously, because our simulated system only comprises around 170 particles, the system will typically be observed to switch between two different configurations, in which either species dominates. 
At $T=20 \epsilon_\circ$ [Fig.~\ref{fig:gcresults}(a)] the populations of each species fluctuates around the same mean value. For $T=10 \epsilon_\circ$ [Fig.~\ref{fig:gcresults}(b)] we observe longer periods of population imbalance, whereas at the lowest temperature $T=5 \epsilon_\circ$ [Fig.~\ref{fig:gcresults}(c)] we observe a species 1-rich phase over the entire duration of the simulation, indicative of phase separation {\cite{emphasis2}}. For comparison we have also plotted the populations for each species for particles obeying Boltzmann statistics ({\it i.e.}, where exchanges are not permitted)
at the lowest temperature.
In this case we find, as expected, that $N_1 \approx N_2$. Thus, demixing in the symmetric mixture is a direct result of Bose statistics~\cite{Boninsegni2012}.
We also fail to see demixing if the simulation is carried out for values of the model parameters $r_s$ or $a$ such that the droplet crystal phase of the single-component Bose system is {\it not} supersolid \cite{Cinti2010}.

To make the concept of phase separation more rigorous we define the \emph{demixing} parameter:
\begin{eqnarray}
D = \langle ( N_1 - N_2 )^2 \rangle / \langle ( N_1 + N_2 )^2 \rangle,
\end{eqnarray}
which is related to the expectation value of the  square of the isospin $S_z$ operator of the underlying SU(2) algebra.
 By construction, when $D = 1$ the system is fully demixed whereas $D=0$ corresponds to a mixed state.  
\\
%Fig.~\ref{fig:gcresults2} shows $D$ as a function of temperature. In order to assess that phase separation ($D > 0$) is concomitant with supersolidity, we show on the same plot the superfluid fraction of the equivalent one-component system, calculated using the winding number estimator. In the limit of zero temperature the superfluid fraction saturates to the value $f_s =  0.40 \pm 0.05$. While the two curves do not have the same shape, as they pertain to different physical effects, the results nonetheless are consistent with the conclusion that the superfluid behaviour of the system leads to demixing.
\indent
Fig.~\ref{fig:gcresults2}(a) shows $D$ as a function of temperature. In order to assess that phase separation ($D > 0$) is concomitant with supersolidity, we show in Fig.~\ref{fig:gcresults2}(b) the superfluid fraction of component 1, denoted $f_s$, calculated using the winding number estimator. For clarity we omit the superfluid fraction for component 2 noting that this quantity is equal for both components to within statistical error. To determine the nature of the transition we have performed simulations for three different system sizes with box areas given by $(L/r_\circ)^2 = 1.92, 7.68, 30.72$ respectively. 
It is clear that both the supersolid and demixing transitions sharpen as the system size increases, consistent with a first order phase transition. Moreover, the plots support the assertion that these transitions occur simultaneously and at non-zero temperature.\\
\indent
It should be noted that the small residual value $D \approx 4\%$ at the highest temperature and smallest system size (and where the system is an insulating droplet crystal)  is a finite size effect. It is indicative of the fact that even above the supersolid transition temperature, individual droplets are preferentially composed of a single species. %, a phenomenon we denote as \emph{local} demixing. 
There exists a free energy cost associated to the {\it tagging} of an individual particle inside a cluster comprising only particles of the same ``color". This free energy cost arises from the fact that a particle that is distinguishable from the others cannot be part of exchange cycles \cite{rmp}.  The tagging of an individual particle has been predicted to cause the expulsion of a single orthohydrogen molecule from the center of a small parahydrogen cluster, {\it i.e.}, another system in which the masses and the interactions are very nearly the same \cite{Mezzacapo2007}. 
This leads to fluctuations in the total populations for each species of the order $1/N_d$ where $N_d$ is the total number of droplets.
\\ \indent
There are in fact three characteristic scales that determine the phase diagram of the symmetric two component system: the thermal de Broglie wavelength $\lambda_{dB} = (2\pi\hbar^2/m k_B T)^{1/2}$, the mean interparticle separation $r_s = 1/\sqrt{\rho}$, and the Rydberg blockade interaction cutoff $a$. When the interaction cutoff exceeds the interparticle separation ($a > r_s$), we expect the system to aggregate into droplets purely through classical potential energy considerations. Moreover, as the temperature of the system is lowered, so that the de Broglie wavelength becomes comparable to (and larger than) the interparticle separation, quantum statistical effects become important. In particular, the world line configurations of the system can involve permutation cycles of more than one particle (of a given bosonic species), this being a manifestation of the exchange symmetry of identical bosons (and being the same mechanism responsible for both superfluidity and Bose-Einstein condensation). There are two regimes for which quantum statistical effects prevail. First, when the de Broglie wavelength is less than or of the order of the interaction cutoff, there exists demixing only within individual droplets. %, attendant with local demixing and superfluidity.
Second, when the de Broglie wavelength is comparable to or larger than the interaction cutoff,  permutation cycles can involve particles in adjacent droplets and droplets that are farther away. It is in this situation that phase separation emerges as well as long range order, so the system becomes supersolid. 
%The energetic mechanism that leads to demixing in both cases is that particles can lower their kinetic energy by exchanging with like particles.%, thereby enhancing their spatial localization. 

%\begin{figure}
%%\begin{center}
%%\begin{minipage}{\columnwidth}
%\includegraphics[scale=1]{rhosDvsT2_final.pdf}
%\caption{(Color online)  Demixing parameter $D$ as a function of temperature. Parameters are $r_s / r_\circ = 0.15$ and $a / r_\circ = 0.27$. Also shown is the superfluid fraction $f_s$ for the equivalent one component system, calculated at two temperatures. Where error bars are absent, statistical errors are smaller than the point size. }
%\label{fig:gcresults2}
%%\end{minipage}%
%%\end{center}
%\end{figure}

\begin{figure}[!t]
\begin{center}
\begin{minipage}{\columnwidth}
  \centering
    \hspace{1cm}(a)\\ 
 \includegraphics[scale=1.1]{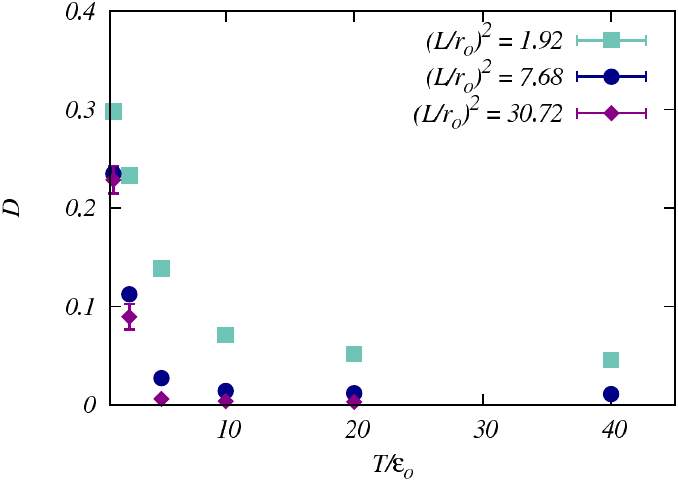}\\
 \vspace{2mm}
  \label{fig:test2}
    %\captionof*{figure}{(c)}\vspace{-2mm}
    \hspace{1cm}(b)\\
  \hspace{-3mm}\includegraphics[scale=1.14]{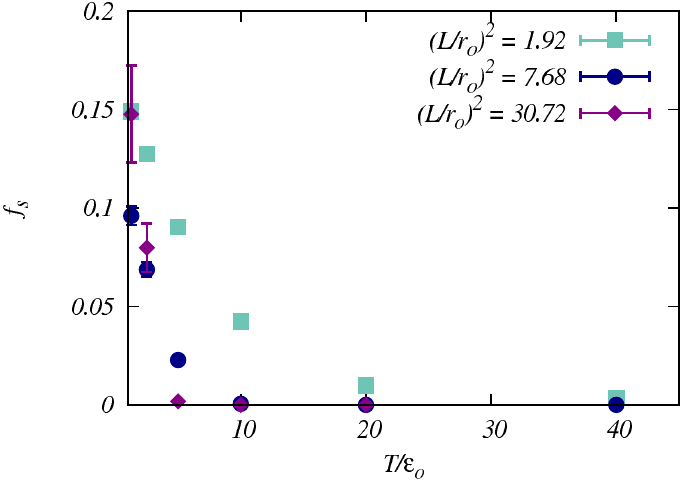}\\
  \vspace{2mm}
  \label{fig:test3}
  %\caption*{\figurename s:~The general caption}
 \caption{(Color online)  (a) Demixing parameter $D$ as a function of temperature.  (b) Corresponding superfluid fraction $f_s$ for component $a$. Parameters are $r_s / r_\circ = 0.15$ and $a / r_\circ = 0.27$. Where error bars are absent, statistical errors are smaller than the point size. }
\label{fig:gcresults2}
\end{minipage}%
\end{center}
\end{figure}

\indent
%The mechanism that leads to demixing is the same as which underlies the supersolid transition in the single-component system, namely tunnelling of particles across adjacent droplets. 
%\color{red} TODO...
The mechanism that leads to demixing can equally be observed in lattice models. The lattice two-component  Bose-Hubbard model (2CBHM), obtained by setting $V(\bm{r} - \bm{r'}) = U \delta(\bm{r} - \bm{r'})$ in Eq.~\ref{eq:ham} and performing a tight binding analysis to the lowest Wannier band in the presence of an external lattice potential, has clear parallels with the free-space supersolid system considered here since each multiply-occupied lattice site acts like a single droplet in the supersolid system.
With equal values of the nearest-neighbor hopping and the on-site inter- and intra-species repulsion, and with a density that is sufficiently high, the same type of phase separation is observed. Each of the two components breaks a U(1) symmetry, and the ground state wavefunction is 
%a Schr{\"o}dinger cat consisting of 
a superposition of the phase separated condensates, leading to spatial demixing. For unit density and strong interactions however, the phase diagram features a phase with counterflow in which only one U(1) symmetry is broken. The existence of such a phase was previously shown in~\cite{Capogrosso2009} for hard-core bosons. In case of SU(2) symmetric interactions and a  phase with counterflow, no superposition takes
place.
Finally, we remark that the difference in ground state wavefunction between a Boltzmann and a Bose system vanishes according to Feynman's theorem~\cite{Feynman1954}. Remixing is hence expected at exactly $T=0$ in our system as well. 
%and this can be understood from the quantum superposition of the cat states.

In summary, we have shown that a symmetric 2-component bosonic crystal with equal inter- and intra-species Rydberg-blockaded interactions  exhibits phase separation, as is evident from a spontaneous symmetry breaking of the particle number for each species, concomitantly with its transition to a supersolid 
phase at low temperature.
%Through the formation of a Schr{\"o}dinger cat in a solid phase, such bosonic systems may 
%thus featuring a hallmark demixing effect when supersolidity sets in. 
Controlling such mixtures during a cooling process and the detection of the demixing are within reach of current technology.
%In this scenario, demixing is attributed to exchange symmetry for identical bosons, which lowers the energy due to both intra- and inter-droplet exchanges. 
 %In contrast at higher temperatures or densities where the system turns insulating, a small ``residual'' demixing is observed, which can be wholly attributed to ``local'' demixing whereby exchanges occur only within individual droplets. 

%\\
%\indent 
%The effect described above could be observed experimentally in a two-component Bosonic system; 

This work was supported in part by the Natural Science and Engineering Research Council of Canada under
research grant G121210893, the Alberta Informatics Circle of Research Excellence, and the University of British Columbia. LP acknowledges financial support from FP7/Marie-Curie Grant No. 321918 ("FDIAGMC") and FP7/ERC Starting Grant No. 306897 ("QUSIMGAS").  The authors would like to thank Fabio Cinti and Aditya Raghavan for useful discussions.


\begin{thebibliography}{99}
\bibitem{emphasis2}
It should be emphasized that during a much longer simulation one expects on average the same number of species 1-rich and species 2-rich configurations due to the symmetry of the Hamiltonian.
\end{thebibliography}
\end{document}